\documentclass[12pt]{article}
 \usepackage{graphicx}
 \usepackage[cp1251]{inputenc} %% Windows
 \usepackage[russian]{babel}   %%

\begin{document}
 \noindent {\footnotesize\it Astronomy Letters, 2015, Vol. 41, No. 3--4, pp. 156--165.}
 \newcommand{\dif}{\textrm{d}}

 \noindent
 \begin{tabular}{llllllllllllllllllllllllllllllllllllllllllllll}
 & & & & & & & & & & & & & & & & & & & & & & & & & & & & & & & & & & & & & \\\hline\hline
 \end{tabular}

 \vskip 1.0cm
 \centerline{\bf Residual HCRF Rotation relative to the Inertial Coordinate System}
 \bigskip
 \centerline{\bf V.V. Bobylev$^{1,2}$}
 \bigskip
 \centerline{\small \it $^1$Pulkovo Astronomical Observatory, St. Petersburg,  Russia}
 \centerline{\small \it $^2$Sobolev Astronomical Institute, St. Petersburg State University, Russia}
 \bigskip
 \bigskip
{\bf Abstract}—VLBI measurements of the absolute proper motions of
23 radio stars have been collected from published data. These are
stars with maser emission, or very young stars, or
asymptotic-giant-branch stars. By comparing these measurements
with the stellar proper motions from the optical catalogs of the
Hipparcos Celestial Reference Frame (HCRF), we have found the
components of the residual rotation vector of this frame relative
to the inertial coordinate system: $(\omega_x,\omega_y,\omega_z) =
(-0.39,-0.51,-1.25)\pm(0.58,0.57,0.56)$ mas yr$^{-1}$. Based on
all the available data, we have determined new values of the
components of the residual rotation vector for the optical
realization of the HCRF relative to the inertial coordinate
system: $(\omega_x,\omega_y,\omega_z) =
(-0.15,+0.24,-0.53)\pm(0.11,0.10,0.13)$ mas yr$^{-1}$.

 %DOI: 10.1134/S1063773715040039

\section*{INTRODUCTION}
The present-day standard astronomical coordinate system, the
International Celestial Reference System (ICRS), has been an
official one since January 1, 1998, by the IAU decision. The
directions of the principal plane and the coordinate origin at the
epoch J2000.0 were fixed in the IAU decision.

This system is realized as a catalog of the positions of 212
compact extragalactic radio sources uniformly distributed over the
entire sky. The catalog was compiled from very long baseline
interferometry (VLBI) observations, which is reflected in Ma et
al.~(1998). This specific realization of the International
Celestial Reference System is designated as ICRF (International
Celestial Reference Frame).

The current version of the catalog, ICRF2, already uses 295 most
reliable radio sources, and the total number of objects in it is
3414 (Ma et al. 2009). The parameters of the third version, ICRF3,
are now being discussed (Jacobs et al. 2014). The selection of
candidates for ICRF3 is planned to be finished by 2018 in order
that the catalog be ready before the completion (by about 2021) of
the GAIA optical space experiment.

The first realization of the ICRS in the optical range was the
Hipparcos (1997) Catalogue, which is usually designated as HCRF
(Hipparcos Celestial Reference Frame). As was shown by Kovalevsky
et al. (1997), the Hipparcos system was tied to the extragalactic
reference frame with an accuracy of $\pm$0.6~mas for the
coordinates at the epoch 1991.25 and with an error of
$\pm$0.25~mas yr$^{-1}$ for the rotation around three axes. The
difficulty of tying Hipparcos to the ICRF lies in the fact that
there are virtually no common objects, because in the optical
range the quasars turned out to be too faint to be directly
observed in the Hipparcos experiment. Therefore, the main
observing program for tying Hipparcos to the ICRF was the program
of VLBI observations of 12 radio stars located near quasars
(Lestrade 1999). The programs to determine the absolute proper
motions of stars relative to galaxies and quasars (Kovalevsky et
al. 1997) also served to check the tying.

The application of various analysis techniques shows that there is
a slight residual rotation of the HCRF relative to the inertial
coordinate system with $\omega_z\approx-0.4$ mas yr$^{-1}$
(Bobylev 2004a, 2004b). To check whether the HCRF is inertial,
VLBI observations of 46 radio stars are being performed (Boboltz
et al. 2003, 2007) in the USA using the ``Very Large Array'' (VLA)
in combination with the Pie Town antenna (New Mexico) at 8.4 GHz.

In this paper, we want to focus our attention on the VLBI
observations of masers and Mira variables that are performed to
study the Galaxy (Reid et al. 2014; Nakagawa et al. 2014). The
VLBI observations of such sources carried out in the last 5--6
years show that the absolute proper motions of radio stars have a
very high accuracy, 0.5--0.05 mas yr$^{-1}$. This accuracy has
been achieved, first, owing to the long interferometer baselines
and, second, owing to the observations at high frequencies,
22.2~GHz and even 43.2~GHz (SiO masers). The goal of this work is
to study the possibility of using such observations to determine
the residual rotation of the HCRF relative to the present-day
realization of the inertial coordinate system.

%%%%%%%%%%%%%%%%%%%%%%%%%%%%%%%%%%%%%%%%%%%%%%
{
\begin{table}[p]                                                %% T-1
\caption[]{\small Characteristics of the radio stars}
\begin{center}
 \small
 \label{t1}
\begin{tabular}{|l|c|c|c|c|c|r|r|r|}\hline

 Star          & HIP/UCAC &  Type &  Frequency, & Program &  Emission &  Ref \\
                 &        &      &    GHz    &        &            &      \\\hline
 T Lep           & HIP 23636 & Mira & 22.2 & VERA &  H$_2$O masers &  (1) \\  %Nakagawa, 2014
 S Crt           & HIP 57917 &  SRb & 22.2 & VERA &  H$_2$O masers &  (2) \\  %NAKAGAWA, 2008
 W Hya           & HIP 67419 &  SRa &  1.6 & NRAO &      OH masers &  (3) \\  %Vlemmings, 2003
 RX Boo          & HIP 70401 &  SRb & 22.2 & VERA &  H$_2$O masers &  (4) \\  %KAMEZAKI, 2012
 S CrB           & HIP 75143 & Mira &  1.6 & NRAO &      OH masers &  (5) \\  %Vlemmings, 2007
 U Her           & HIP 80488 & Mira &  1.6 & NRAO &      OH masers &  (5) \\  %Vlemmings, 2007
 RR Aql          & HIP 98220 & Mira &  1.6 & NRAO &      OH masers &  (5) \\  %Vlemmings, 2007
 R Aqr           & HIP117054 & Mira & 43.2 & VERA &     SiO masers &  (6) \\  %Min, 2014
 R Cas           & HIP118188 & Mira &  1.6 & NRAO &      OH masers &  (3) \\  %Vlemmings, 2003
 SY Scl          & UCAC4     & Mira & 22.2 & VERA &  H$_2$O masers &  (7) \\  %Nyu, 2011
 UX Cyg          & UCAC4     & Mira & 22.2 & NRAO &  H$_2$O masers &  (8) \\  %Kurayama, 2005
 SS Cyg          & UCAC4   & D. Nova  &  8.4 & NRAO &    continuum &  (9) \\  %Miller-Jones, 2013
 IM Peg          & HIP112997 & RS CVn &  8.4 & VLBI &    continuum & (10) \\  %Ratner, 2012
 S Per           & HIP 11093 & SRc   & 22.2 & NRAO &  H$_2$O masers & (11) \\ %Asaki, 2010
 V773 Tau        & HIP 19762 & T Tau &  8.4 & NRAO &      continuum & (12) \\ %Torres, 2012,
 HDE 283572      & HIP 20388 & T Tau &  8.4 & NRAO &      continuum & (13) \\ %Torres, 2007,
 T Tau N         & HIP 20390 & T Tau &  8.4 & NRAO &      continuum & (14) \\ %Loinard,2007,
 LSI +61 303     & HIP 12469 & XMXRB &  8.4 & VLBI &      continuum & (15) \\ %Dhawan, 2006
 Cyg X-1         & HIP 98298 & XMXRB &  8.4 & NRAO &      continuum & (16) \\ %Reid, 2011
 Cyg OB2\#5      & HIP101341 &  EB   &  8.4 & NRAO &      continuum & (17) \\ %Dzib, 2013
 IRAS 22480+6002 & UCAC4     &  ---  & 22.2 & VERA &  H$_2$O masers & (18) \\ %Imai,   2012
 PZ Cas          & HIP117078 &  SRa  & 22.2 & VERA &  H$_2$O masers & (19) \\ %Kusuno, 2013
$\theta^1$ Ori A & UCAC4     &  ---  &  8.4 & NRAO &      continuum & (20) \\ %Menten, 2007
 \hline
 \end{tabular}
 \end{center}
 \small\baselineskip=1.0ex\protect
SR---semiregular pulsating stars; RS---eruptive variables of the
RS Canum Venaticorum type; EB---eclipsing binaries;
XMXRB---high-mass X-ray binaries; D. Nova---dwarf novae.

(1): Nakagawa et al. (2014), (2): Nakagawa et al. (2008), (3):
Vlemmings et al. (2003), (4): Kamezaki et al. (2012), (5):
Vlemmings and van Langevelde (2007), (6): Min et al. (2014), (7):
Nyu et al. (2011), (8): Kurayama et al. (2005), (9): Miller-Jones
et al. (2013), (10): Ratner et al. (2012), (11): Asaki et al.
(2010), (12): Torres et al. (2012), (13): Torres et al. (2007),
(14): Loinard et al. (2007), (15): Dhawan et al. (2006), (16):
Reid et al. (2011), (17): Dzib et al. (2013), (18): Imai et al.
(2012), (19): Kusuno et al. (2013), (20): Menten et al. (2007).
\end{table}
}
%%%%%%%%%%%%%%%%%%%%%%%%%%%%%%%%%%%%%%%%%%%%%%
%%%%%%%%%%%%%%%%%%%%%%%%%%%%%%%%%%%%%%%%%%%%%%
{
\begin{table}[t]                                                %% T-2
\caption[]{\small Proper motions of the radio stars}
\begin{center}
 \small
 \label{t2}
\begin{tabular}{|l|r|r|r|r|r|r|r|r|}\hline
                 & \multicolumn{2}{c|} {VLBI}&\multicolumn{2}{c|} {HIPPARCOS-2007} \\\hline
 Star            & $\mu_\alpha\cos\delta\pm e,$ & $\mu_\delta\pm e,$ & $\mu_\alpha\cos\delta\pm e,$ & $\mu_\delta\pm e,$ \\
                 &   mas yr$^{-1}$ & mas yr$^{-1}$ & mas yr$^{-1}$ &  mas yr$^{-1}$   \\\hline
 T Lep           & $ 14.60\pm0.50$ & $-35.43\pm0.70$ & $ 11.43\pm0.98$ & $-33.34\pm0.86$ \\
 S Crt           & $ -3.17\pm0.22$ & $ -5.41\pm0.22$ & $ -3.37\pm1.00$ & $ -4.68\pm0.75$ \\
 W Hya           & $-44.24\pm2.04$ & $-55.28\pm2.98$ & $-49.31\pm1.48$ & $-59.71\pm1.04$ \\
 RX Boo          & $ 24.55\pm1.06$ & $-49.67\pm2.38$ & $ 21.21\pm0.50$ & $-48.79\pm0.46$ \\
 S CrB           & $ -9.06\pm0.23$ & $-12.52\pm0.29$ & $ -7.73\pm0.57$ & $-13.03\pm1.02$ \\
 U Her           & $-14.98\pm0.29$ & $ -9.23\pm0.32$ & $-15.94\pm0.61$ & $-11.03\pm0.69$ \\
 RR Aql          & $-25.11\pm0.74$ & $-49.82\pm0.54$ & $-25.12\pm4.46$ & $-49.14\pm3.23$ \\
 R Aqr           & $ 37.13\pm0.47$ & $-28.62\pm0.44$ & $ 33.00\pm1.53$ & $-25.74\pm1.30$ \\
 R Cas           & $ 80.52\pm2.35$ & $ 17.10\pm1.75$ & $ 85.52\pm0.75$ & $ 17.49\pm0.72$ \\
 SY Scl          & $  5.57\pm0.04$ & $ -7.32\pm0.12$ & $  4.60\pm0.90$ & $ -7.80\pm0.90$ \\
 UX Cyg          & $ -6.91\pm0.75$ & $-12.52\pm1.57$ & $ -8.90\pm1.60$ & $ -9.80\pm2.50$ \\
 SS Cyg          & $112.42\pm0.07$ & $ 33.38\pm0.07$ & $113.20\pm0.90$ & $ 33.40\pm1.00$ \\
 IM Peg          & $-20.83\pm0.09$ & $-27.27\pm0.09$ & $-20.73\pm0.28$ & $-27.75\pm0.27$ \\
 S Per           & $ -0.49\pm0.35$ & $ -1.19\pm0.33$ & $ -2.70\pm2.20$ & $ -0.29\pm1.65$ \\
 V773 Tau        & $  8.30\pm0.50$ & $-23.60\pm0.50$ & $  4.11\pm2.74$ & $-24.48\pm1.88$ \\
 HDE 283572      & $  8.88\pm0.06$ & $-26.60\pm0.10$ & $  6.84\pm1.64$ & $-27.15\pm1.12$ \\
 T Tau N         & $ 12.35\pm0.04$ & $-12.80\pm0.05$ & $ 15.51\pm1.93$ & $-13.67\pm1.64$ \\
 LSI +61 303     & $ -0.30\pm0.07$ & $ -0.26\pm0.05$ & $  0.27\pm2.91$ & $  2.38\pm2.52$ \\
 Cyg X-1         & $ -3.78\pm0.06$ & $ -6.40\pm0.12$ & $ -3.37\pm0.75$ & $ -7.15\pm0.86$ \\
 Cyg OB2\#5      & $ -1.64\pm0.98$ & $ -7.16\pm1.28$ & $ -2.14\pm1.23$ & $ -2.48\pm1.14$ \\
 IRAS 22480+6002 & $ -2.58\pm0.33$ & $ -1.91\pm0.17$ & $ -2.60\pm1.90$ & $ -2.10\pm6.10$ \\
 PZ Cas          & $ -3.70\pm0.20$ & $ -2.00\pm0.30$ & $ -4.15\pm0.85$ & $ -3.55\pm0.81$ \\
 $\theta^1$ Ori A & $  4.82\pm0.09$ & $ -1.54\pm0.18$ & $-0.90\pm2.50$ & $  0.10\pm2.50$ \\\hline
 \end{tabular}
 \end{center}
 \end{table} }
%%%%%%%%%%%%%%%%%%%%%%%%%%%%%%%%%%%%%%%%%%%%%%%%

\section*{VLBI DATA}
One of the projects to measure the trigonometric parallaxes and
proper motions is the Japanese VERA (VLBI Exploration of Radio
Astrometry) project devoted to the observations of H$_2$O masers
at 22.2~GHz (Hirota et al. 2007) and SiO masers at 43.2~GHz (Kim
et al. 2008).

Methanol (CH$_3$OH, 6.7 and 12.2 GHz) and H$_2$O masers are being
observed in the USA on the VLBA (Reid et al. 2014). Similar
observations are also being carried out within the framework of
the European VLBI network (Rygl et al. 2010). These two programs
enter into the general BeSSeL\footnote
{http://www3.mpifr-bonn.mpg.de/staff/abrunthaler/BeSSeL/index.shtml}
(Bar and Spiral Structure Legacy Survey) project (Brunthaler et
al. 2011).

For the same purpose, the VLBI observations of radio stars are
being carried out in continuum at 8.4~GHz (Torres et al. 2007;
Dzib et al. 2012). The radio sources located in the local arm
(Orion arm) that are associated with young low-mass protostars are
being observed within the framework of this program.

Mira variables at the asymptotic-giant-branch (AGB) stage are
being observed in the VERA program. Such stars manifest themselves
in the radio band as masers (Nakagawa et al. 2014). Thus, our
sample includes young massive supergiants, intermediate-mass
giants, and low-mass T Tauri stars.

An important peculiarity of all the listed observations is that
the stellar positions have always been determined using two or
three extragalactic sources, i.e., the parallaxes and proper
motions of these radio stars are absolute.

Table 1 presents some characteristics of the radio stars. Columns
1, 2, 3, 4, 5, 6, and 7 give, respectively, the star name, the
number in the Hipparcos Catalogue, the type of variability, the
observation frequency, the name of the observing program or
observatory, the type of emission characterizing the VLBI
observations (for example, they were performed in maser lines or
in continuum), and references. Note that the Hipparcos proper
motions are very unreliable for the massive O star $\theta^1$
Ori\,A (HIP 26220), a member of the famous Orion Trapezium;
therefore, we used the data from the UCAC4 (2012) catalog. It can
be seen from the second column of Table~1 that there are four more
stars (absent in the Hipparcos Catalogue) with the UCAC4 proper
motions.

We did not include the red supergiant VY\,CMa (HIP 35793) in our
list, for which high-accuracy VLBI observations are available
(Zhang et al. 2012). For this star, the absolute values of the
``Hipparcos minus VLBI'' proper motion differences are about 8~mas
yr$^{-1}$ in each of the coordinates. Such a large difference
stems from the fact that slightly different parts of the extended
asymmetric envelope of this star are observed in the optical and
radio bands.

Note that long-term observations in maser lines are difficult to
carry out, because the maser spots are highly unstable; therefore,
the periods of observations for such stars are 1.5--2 years (the
minimum period required to determine the annual parallax). The
situation is different for continuum observations. For example,
the period of observations for IM\,Peg was about 8~years (Ratner
et al. 2012).

Table 2 presents the proper motions of the radio stars measured by
the VLBI technique (according to the references in Table 1) and
the proper motions of these stars from the version of the
Hipparcos Catalogue revised by van Leeuwen (2007). Van Leeuwen
(2007) showed that the new version of the Catalogue completely
reproduces the previous system, i.e., it has no residual rotation
relative to the Hipparcos-1997 version, while it excels
considerably the Hipparcos-1997 version in terms of random errors
(especially in the region of bright stars). It is of interest to
check this using the available material.

Figure 1 presents the ``HIP-1997 minus VLBI'' and ``HIP-2007 minus
VLBI'' stellar proper motion differences. It can be clearly seen
from our comparison of the graphs that the dispersion of the
``HIP-1997 minus VLBI'' differences exceeds considerably that of
the ``HIP-2007 minus VLBI'' differences. For some of the stars,
the absolute values of the ``HIP-1997 minus VLBI'' differences in
right ascension exceed 6 mas yr$^{-1}$.

In Fig.~2, the ``HIP-2007 minus VLBI'' stellar proper motions are
plotted against the equatorial coordinates $\alpha$ and $\delta$.
It can be seen from the figure that a wave in $\alpha$ and a
slight trend in $\delta$ are noticeable in the differences
$\Delta\mu_\alpha\cos\delta$, while the differences
$\Delta\mu_\delta$ are distributed quite symmetrically relative to
the horizontal axis.

The following coupling equations can be used to determine the
three angular velocities of mutual rotation of the two systems
$\omega_x,\omega_y,\omega_z:$
 \begin{equation}
 \begin{array}{lll}
 \Delta\mu_\alpha\cos\delta=
 - \omega_x\cos\alpha\sin\delta - \omega_y\sin\alpha\sin\delta + \omega_z\cos\delta, \\
 \Delta\mu_\delta= \omega_x\sin\alpha - \omega_y\cos\alpha,
 \label{HIP-ABS}
 \end{array}
 \end{equation}
where the ``Hipparcos minus VLBI'' differences are on the
left-hand sides of the equations. Having solved the system of
conditional equations (1) by the least squares method for the
``HIP-1997 minus VLBI'' difference, we obtained the following
rotation components (in mas yr$^{-1}$):
 \begin{equation}
 \begin{array}{lll}
 \omega_x=-0.85\pm0.63, \\
 \omega_y=+0.05\pm0.61, \\
 \omega_z=-1.33\pm0.61,
 \label{HIP-97}
 \end{array}
 \end{equation}
where the error per unit weight, which characterizes the
dispersion of the residuals, is $\sigma_0=2.37$ mas yr$^{-1}$. For
the ``HIP-2007 minus VLBI'' difference, we obtained the following
rotation components (in mas yr$^{-1}$):
 \begin{equation}
 \begin{array}{lll}
 \omega_x=-0.39\pm0.58, \\
 \omega_y=-0.51\pm0.57, \\
 \omega_z=-1.25\pm0.56,
 \label{HIP-2007}
 \end{array}
 \end{equation}
where the error per unit weight is $\sigma_0=2.19$ mas yr$^{-1}$.
We see that it is more advantageous to use the HIP-2007 version.

%%%%%%%%%%%%%%%%%%%%%%%%%%%%%%%%%%%%%%%% f.1:
 \begin{figure}[t]
 {\begin{center}
 \includegraphics[width=140mm]{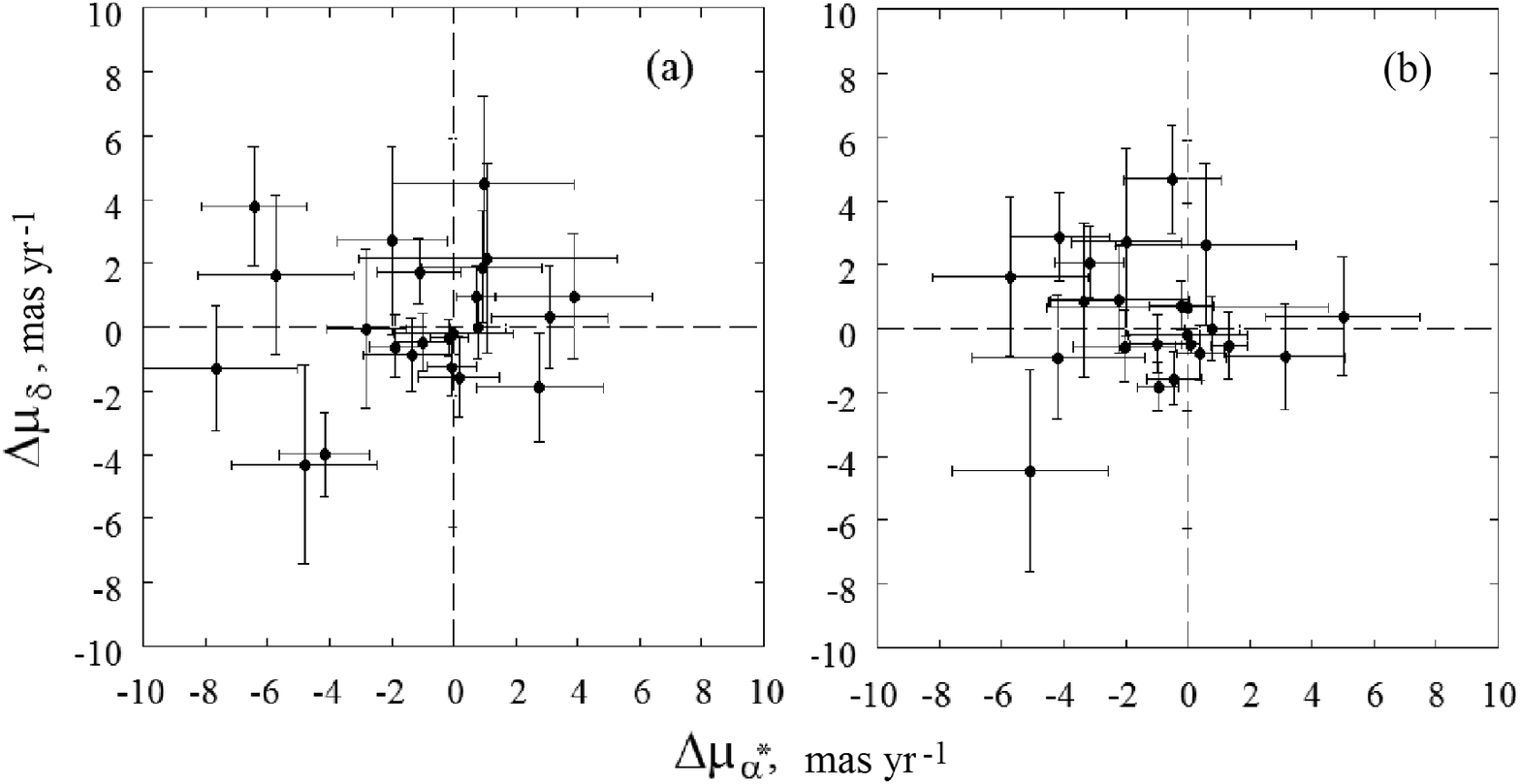}
 \caption{
 (a) ``HIP-1997 minus VLBI'' stellar proper motion differences and
 (b) ``HIP-2007 minus VLBI'' stellar proper motion differences;
     $\Delta\mu_{\alpha^*}=\Delta\mu_\alpha\cos\delta$.}
 \label{fig1}
 \end{center} }
 \end{figure}
%%%%%%%%%%%%%%%%%%%%%%%%%%%%%%%%%%%%%%%%%%%%%%%
%%%%%%%%%%%%%%%%%%%%%%%%%%%%%%%%%%%%%%%% f.2:
 \begin{figure}[t]
 {\begin{center}
 \includegraphics[width=150mm]{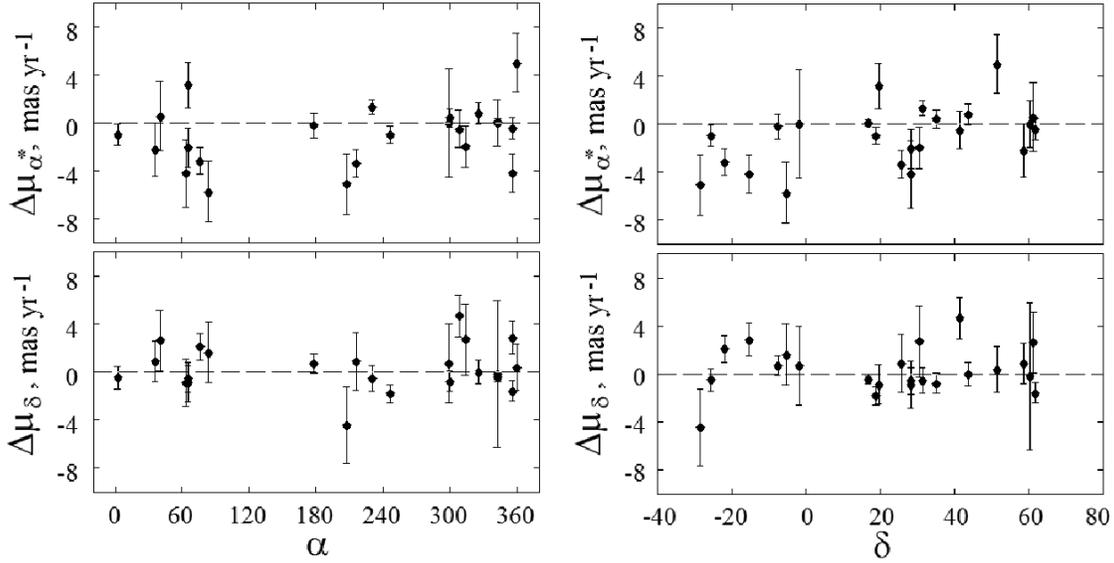}
 \caption{``HIP-1997 minus VLBI'' stellar proper motion differences versus
 equatorial coordinates $\alpha$ and $\delta$;
 $\Delta\mu_{\alpha^*}=\Delta\mu_\alpha\cos\delta$.}
 \label{fig2}
 \end{center} }
 \end{figure}
%%%%%%%%%%%%%%%%%%%%%%%%%%%%%%%%%%%%%%%%%%%%%%%

\section*{HCRF INERTIALITY}
Table~3 presents almost all of the results of comparing the
individual programs with the HCRF catalogs known to date. The
first seven rows of Table~3 give the results used by Kovalevsky et
al. (1997) to calibrate Hipparcos and to estimate its residual
rotation relative to the system of extragalactic sources.

(1) The ``VLBI-1999'' solution was obtained by comparing the
absolute proper motions of twelve radio stars with the Hipparcos
Catalogue. A detailed description of the observations can be found
in Lestrade et al. (1999). In total, 21 antennas of various
diameters in the USA and Europe were involved in the observations.
The period of observations for each star was from 2 to 11 years.
The observations were performed at 5.0 and 8.4~GHz. The
coordinates, parallaxes, proper motions, and even accelerations in
the proper motions of stars were determined. According to the
estimates by Lestrade et al. (1999), the error in the stellar
proper motion for some of the stars was about 0.05 mas yr$^{-1}$.

(2) As the ``NPM1'' solution, we used the results of comparing the
stellar proper motions from the NPM1 (Klemola et al. 1994) and
Hipparcos Catalogues by the Heidelberg team. We took the
parameters of this solution from Table~2 in Kovalevsky et al.
(1997). This solution was obtained in the range of magnitudes
 10$^m$.5--11$^m$.5, where (Fig.~1 in Platais et al. (1998a)) the
``HIP minus NPM1'' stellar proper motion differences have a
``horizontal'' character, near zero. In our opinion, the proper
motions from the NPM1 Catalogue in this magnitude range are free
to the greatest extent from the magnitude equation, which is
considerable in this Catalogue.

(3) The ``KIEV'' solution. The stellar proper motions from
theGPM1(Rybka and Yatsenko 1997) and Hipparcos Catalogues were
compared by Kislyuk et al. (1997).

(4) The ``POTSDAM'' solution was taken from Hirte et al. (1996).

(5) The ``BONN'' solution was obtained by performing the Bonn
program, which is reflected in Geffert et al. (1997) and Tucholke
et al. (1997).

(6) The ``EOP'' solution was obtained as a result of the analysis
of the Earth Orientation Parameters (EOPs) by Vondr\'ak et al.
(1997). Only two orientation parameters, $\omega_x$ and
$\omega_y,$ are determined in this method.

(7) The ``HST'' solution was obtained by Hemenway et al. (1997)
using the Hubble Space Telescope (HST). Note that this solution
has virtually no influence on the calculation of the weighted
average because of the large random errors in the comparison
parameters.

(8) The ``SPM2'' solution was obtained by Zhu (2001) by comparing
the stellar proper motions from the SPM2 (Platais et al. 1998b)
and Hipparcos Catalogues.

(9) The ``PUL2'' solution was found by comparing the PUL2 Pulkovo
photographic catalog (Bobylev et al. 2004) and the Hipparcos
Catalogue.

(10) The ``XPM'' solution is based on the Khar'kov catalog of
absolute stellar proper motions, XPM (Fedorov et al. 2009). It was
absolutized using $\approx$1.5 million galaxies from the 2MASS
catalog of extended sources (Skrutskie et al. 2006). Thus, the XPM
catalog is an independent realization of the inertial coordinate
system. The proper motions from the XPM and UCAC2 (Zacharias et
al. 2004) catalogs were compared in Bobylev et al. (2010), where
the parameters $\omega_x$, $\omega_y$, and $\omega_z$ were
calculated using $\approx$1~million stars.

(11) The ``MINOR PLANETS'' solution. Chernetenko et al. (2008)
estimated the orientation parameters of the Hipparcos system
relative to the coordinate systems of the DE403 and DE405
ephemerides by analyzing a long-term series of asteroid
observations. This result leads to the conclusion that either the
DE403 and DE405 dynamical theories need an improvement or the
Hipparcos system needs a correction. We reduced the weight of this
solution by half because of the possible contribution from the
inaccuracy of the DE403 and DE405 dynamical theories.

(12) The ``VLA + PT-2007'' solution was obtained by Boboltz et al.
(2007) by analyzing the positions and proper motions of 46 radio
stars. The VLBI observations were performed in the USA using the
``Very Large Array'' (VLA) in combination with the Pie Town
antenna (New Mexico) at 8.4~GHz. In comparison with the
observations by Lestrade et al. (1999), these observations have a
considerably lower resolution in positional observations, because
the interferometer size is smaller. On average, the positional
errors are 13~mas in right ascension and 16 mas in declination
(Boboltz et al. 2003). To achieve a high accuracy of determining
the proper motions, they used long series of observations, about
20 years for each star. Only the coordinates and proper motions of
radio stars were the quantities being determined. According to the
estimates by Boboltz et al. (2007), the error in the stellar
proper motion in this program is, on average, $\approx$1.7 mas
yr$^{-1}$ in each coordinate.

(13) The search for the ``VLBI-2014'' solution is described in the
first part of this paper. Note that the three solutions under
consideration based on the VLBI observations of radio stars,
``VLBI-1999'', ``VLA + PT-2007'', and ``VLBI-2014'', have
significant differences in observing techniques; therefore, we
consider them as independent solutions.

The weight assigned to each comparison catalog is inversely
proportional to the square of the mean error $e_\omega$ in the
corresponding quantities $(\omega_x,\omega_y,\omega_z)$ and was
calculated from the formula
 \begin{equation}
 P_i={e_{kiev}}^2/{{e_i}^2}, \quad i=1,...,13.
 \end{equation}
Equations of the form (1), where the ``Hipparcos minus Catalogue''
differences are on the left-hand sides, can serve to determine
$\omega_x$, $\omega_y$, and $\omega_z$. Lindegren and Kovalevsky
(1995) proposed a slightly different form of these equations:
 \begin{equation}
 \begin{array}{lll}
 \Delta\mu_\alpha\cos\delta=
 \omega_x\cos\alpha\sin\delta+\omega_y\sin\alpha\sin\delta - \omega_z\cos\delta, \\
 \Delta\mu_\delta=
 -\omega_x\sin\alpha + \omega_y\cos\alpha,
 \label{ABS-HIP}
 \end{array}
 \end{equation}
where the ``Catalogue minus Hipparcos'' differences are on the
left-hand sides of the equations. Zhu~(2001) and Boboltz et al.
(2007) published the parameters that they determined with a change
in the form of either the left-hand sides or the righthand sides
of Eqs.~(1) or (5). In these two cases, we brought the signs of
the quoted parameters to the necessary standard form.

The last rows of Table~3 present ``Average~1'' that was calculated
as a simple average but without using the HST solution and
``Average~2'' that was calculated from all data as a weighted mean
and is the main result of our analysis. It is important to note
that the rotation component $\omega_z =-0.53\pm0.13$ mas yr$^{-1}$
in the solution obtained differs significantly from zero.

%%%%%%%%%%%%%%%%%%%%%%%%%%%%%%%%%%%%%%%%%%%%%%
{
\begin{table}[t]                                                %% T~3
\caption[]{\small Components of the residual rotation vector of
the optical realization of the ICRS/Hipparcos system relative to
the inertial coordinate system
 }
\begin{center}
 \label{t3}
\begin{tabular}{|l|c|c|c|c|c|c|c|c|c|c|c|c|c|}\hline

 Method   & $N_\star$ & $N_{\hbox{\tiny area}}$
 & $\omega_x,$~mas yr$^{-1}$
 & $\omega_y,$~mas yr$^{-1}$
 & $\omega_z,$~mas yr$^{-1}$ \\\hline

 VLBI-1999     &    12 &      & $-0.16\pm0.30$ & $-0.17\pm0.26$ & $-0.33\pm0.30$ \\
 NPM1          &  2616 &  899 & $-0.76\pm0.25$ & $+0.17\pm0.20$ & $-0.85\pm0.20$ \\
 Kiev          &   415 &  154 & $-0.27\pm0.80$ & $+0.15\pm0.60$ & $-1.07\pm0.80$ \\
 Potsdam       &   256 &   24 & $+0.22\pm0.52$ & $+0.43\pm0.50$ & $+0.13\pm0.48$ \\
 Bonn          &    88 &   13 & $+0.16\pm0.34$ & $-0.32\pm0.25$ & $+0.17\pm0.33$ \\
 EOP           &       &      & $-0.93\pm0.28$ & $-0.32\pm0.28$ & --- \\
 HST           &    78 &      & $-1.60\pm2.87$ & $-1.92\pm1.54$ & $+2.26\pm3.42$ \\
 SPM2          &  9356 &  156 & $+0.10\pm0.17$ & $+0.48\pm0.14$ & $-0.17\pm0.15$ \\
 PUL2          &  1004 &  147 & $-0.98\pm0.47$ & $-0.03\pm0.38$ & $-1.66\pm0.42$ \\
 XPM   & $1\times10^6$ & 1431 & $-0.06\pm0.15$ & $+0.17\pm0.14$ & $-0.84\pm0.14$ \\
 Minor Planets &   116 &      & $+0.12\pm0.08$ & $+0.66\pm0.09$ & $-0.56\pm0.16$ \\
 VLA+PT-2007   &    46 &      & $-0.55\pm0.34$ & $-0.02\pm0.36$ & $+0.41\pm0.37$ \\
 VLBI-2014     &    23 &      & $-0.39\pm0.58$ & $-0.51\pm0.57$ & $-1.25\pm0.56$ \\\hline
 Average 1 &          &      & $-0.29\pm0.12$ & $+0.06\pm0.10$ & $-0.55\pm0.20$ \\
 Average 2 &          &      & $-0.15\pm0.11$ & $+0.24\pm0.10$ & $-0.53\pm0.13$ \\\hline

\end{tabular}
\end{center}
 \small\baselineskip=1.0ex\protect
 $N_\star$ is the number of stars/asteroids,
 $N_{\hbox{\tiny area}}$  is the number of areas on the celestial sphere, average~1
is a simple average (without HST); average~2 is a weighted
average.
\end{table}
}
%%%%%%%%%%%%%%%%%%%%%%%%%%%%%%%%%%%%%%%%%%%%%%

\section*{DISCUSSION}
By comparing the XPM stars with catalogs extending the HCRF to
faint magnitudes, such as PPMXL (Roeser et al. 2010), UCAC3
(Zacharias et al. 2009), Tycho-2 (Hog et al. 2000), and XC1
(Fedorov and Myznikov 2006), Fedorov et al. (2011) showed that the
two rotation components $\omega_x$ and $\omega_y$ do not differ
significantly from zero, while the $Z$ rotation component has a
measurable value, $\omega_z = -1.8\pm0.16$ mas yr$^{-1}$.

Note the result by Bobylev and Khovritchev (2011) obtained from
the ``proper motions'' of about 8000 galaxies from the UCAC3
catalog (Zacharias et al. 2009). These authors found the residual
rotation of the HCRF with respect to the inertial coordinate
system $(\Omega_x,\Omega_y,\Omega_z) =
(0.58,-1.02,-0.59)\pm(0.15,0.15,0.17)$ mas yr$^{-1}$ in Galactic
coordinates. The corresponding rotation components around the
equatorial axes (in mas yr$^{-1}$) are
 \begin{equation}
 \begin{array}{lll}
 \omega_x=-0.02\pm0.15, \\
 \omega_y=+0.06\pm0.15, \\
 \omega_z=-1.31\pm0.17.
 \label{HIP-UCAC3-eqvatorial}
 \end{array}
 \end{equation}
Fedorov et al. (2011), Wu et al. (2011), and Grabowski et al.
(2014) showed the ``proper motions'' of galaxies and quasars from
the PPMXL catalogs to have measurable values. The proper motions
averaged over all galaxies and quasars are $\approx-2$ mas
yr$^{-1}$ in each of the coordinates ($\mu_\alpha\cos\delta$ and
$\mu_\delta$). Wu et al. (2011) and Grabowski et al. (2014) did
not determine the global rotation parameters, but they proposed to
perform an additional absolute calibration of the stellar proper
motions from the PPMXL catalog using a table of corrections.

Another interesting example of allowance for the additional
absolute calibration of the stellar proper motions from the PPMXL
catalog using galaxies is given in L\'opez-Corredoira et al.
(2014). These authors studied the parameters of such a large-scale
phenomenon as the Galactic warp using red giant clump stars. It
can be clearly seen from Fig.~3 in the cited paper that the
corrections to the vertical stellar velocities reach about 70~km
s$^{-1}$ at a heliocentric distance of about 8~kpc toward the
Galactic anticenter. In other words, this distribution can be
interpreted as the existence of residual rotation of the PPMXL
stars around the Galactic $Y$ axis with an angular velocity
$\Omega_y\approx-1.5$ mas yr$^{-1}$, in good agreement with the
result of analyzing the PPMXL catalog obtained by Fedorov et al.
(2011).

All of the aforesaid leads us to conclude that although the values
of $\Omega_z$ that we derived from radio stars in solutions~(2)
and (3) have fairly large errors, they are in agreement with the
results of the analysis of large catalogs with galaxies presented
above, in particular, with the result~(6) and the result from
Fedorov et al. (2011).

Having analyzed the individual solutions but using a smaller
amount of data, Bobylev (2010) found the following rotation
components:
 $(\omega_x,\omega_y,\omega_z)
=(-0.11,+0.24,-0.52)\pm(0.14,0.10,0.16)$ mas yr$^{-1}$. The
advantage of the result of this study obtained in a similar way
(``Average~2'' in Table~3) is that the random errors in the
parameters decreased. Thus, each result of the individual
comparison of catalogs is of great importance for the solution of
our problem.

\section*{CONCLUSIONS}
We collected highly accurate VLBI determinations of the absolute
proper motions of 23 radio stars from published data. The
observations of the parallaxes and proper motions of these stars
have been performed by various scientific teams in the last five
or six years. The stars have a different evolutionary status. Some
of them are very young stars with maser emission (H$_2$O and
CH$_3$OH masers). AGB stars observed as OH, H$_2$O, and SiO masers
constitute the other part of the sample. Several stars were
observed in continuum.

By comparing these measurements with the proper motions from the
HCRF catalogs, we found the components of the residual rotation
vector of this system with respect to the present-day realization
of the inertial coordinate system, namely:
 $(\omega_x,\omega_y,\omega_z)=(-0.39,-0.51,-1.25)\pm(0.58,0.57,0.56)$ mas yr$^{-1}$.
This estimate is a completely new result of the individual
comparison of the sample of radio stars with the Hipparcos
Catalogue.

Based on all the available individual results, we determined new,
most probable values of the components of the residual rotation
vector for the optical realization of the HCRF with respect to the
inertial coordinate system:
 $(\omega_x,\omega_y,\omega_z)=(-0.15,+0.24,-0.53)\pm(0.11,0.10,0.13)$ mas yr$^{-1}$.

\section*{ACKNOWLEDGMENTS}
This work was supported by the ``Nonstationary Phenomena in
Objects of the Universe'' Program P--21 of the Presidium of the
Russian Academy of Sciences. The SIMBAD search database was very
helpful.

 \newpage
 \bigskip{REFERENCES}
 \bigskip
 {\small

1. Y. Asaki, S. Deguchi, H. Imai, K. Hachisuka, M. Miyoshi, and M.
Honma, Astrophys. J. 721, 267 (2010).

2. D.A. Boboltz, A.L. Fey, K.J. Johnston, M.J. Claussen, C. de
Vegt, N. Zacharias, and R.A.~Gaume, Astron. J. 126, 484 (2003).

3. D.A. Boboltz, A.L. Fey, W.K. Puatua, N. Zacharias, M.J.
Claussen, K.J. Johnston, and R.A.~Gaume, Astron. J. 133, 906
(2007).

 4. V.V. Bobylev, N.M. Bronnikova, and N.A. Shakht, Astron. Lett. 30, 469 (2004).

 5. V.V. Bobylev, Astron. Lett. 30, 289 (2004a).

 6. V.V. Bobylev, Astron. Lett. 30, 848 (2004b).

 7. V.V. Bobylev, P.N. Fedorov, A.T. Bajkova, and V.S. Akhmetov, Astron. Lett. 36, 506 (2010).

 8. V.V. Bobylev, Astron. Lett. 36, 634 (2010).

 9. V.V. Bobylev and M.Yu. Khovritchev, MNRAS 417, 1952 (2011).

10. A. Brunthaler, M.J. Reid, K.M. Menten, X.-W. Zheng, A.
Bartkiewicz, Y.K. Choi, T.~Dame, K. Hachisuka, K. Immer, G.
Moellenbrock, et al., Astron. Nach. 332, 461 (2011).

11. Yu.A. Chernetenko, Astron. Lett. 34, 266 (2008).

12. V. Dhawan, A. Mioduszewski, and M. Rupen, in Proceedings of
the 6th Microquasar Workshop: Microquasars and Beyond, September
18--22, 2006, Como, Italy (2006), p. 52.1.

13. S.A. Dzib, L.F. Rodriguez, L. Loinard, A.J. Mioduszewski, G.N.
Ortiz-Le\'on, and A.T.~Araudo, Astrophys. J. 763, 139 (2013).

 14. P.N. Fedorov and A.A. Myznikov, Kinem. Fiz. Nebesn. Tel 22, 309 (2006).

 15. P.N. Fedorov, A.A. Myznikov, and V.S. Akhmetov, Mon. Not. R. Astron. Soc. 393, 133 (2009).

 16. P.N. Fedorov, V.S. Akhmetov, and V.V. Bobylev, Mon. Not. R. Astron. Soc. 416, 403 (2011).

17. M. Geffert, A.R. Klemola, M. Hiesgen, and J. Schmoll, Astron.
Astrophys. 124, 157 (1997).

18. K. Grabowski, J.L. Carlin, H.J. Newberg, T.C. Beers, L. Chen,
L. Deng, C.J. Grillmair, P. Guhathakurta, J. Hou, S. L\'epine, et
al., arXiv:1409.2890 (2014).

19. P.D. Hemenway, R.L. Duncombe, E.P. Bozyan, A.M. Lalich, A.N.
Argue, O.G. Franz, B.~McArthur, E. Nelan, D. Taylor, et al.,
Astron. J. 114, 2796 (1997).

20. T. Hirota, T. Bushimata, Y.K. Choi, M. Honma, H. Imai,
K.~Iwadate, T. Jike, S.~Kameno, O.~Kameya, R. Kamohara, et al.,
Publ. Astron. Soc. Jpn. 59, 897 (2007).

21. S. Hirte, E. Schilbach, and R.-D. Scholz, Astron. Astrophys.
Suppl. Ser. 126, 31 (1996).

22. E. Hog, C. Fabricius, V.V. Makarov, S. Urban, T. Corbin, G.
Wycoff, U. Bastian, P.~Schwekendiek, and A. Wicenec, Astron.
Astrophys. 355, L27 (2000).

23. H. Imai, N. Sakai, H. Nakanishi, H. Sakanoue, M. Honma, and T.
Miyaji, Publ. Astron. Soc. Jpn. 64, 142 (2012).

24. C. Jacobs, F. Arias, D. Boboltz, J. Boehm, S. Bolotin, G.
Bourda, P. Charlot, A. de Witt, A. Fey, et al., in Proceedings of
the 223rd American Astronomical Society Meeting, Washington, DC,
January 5--9, 2014 (2014), Abstract No. 251.25.

25. T. Kamezaki, A. Nakagawa, T. Omodaka, T. Kurayama, H. Imai, D.
Tafoya, M. Matsui, and Y. Nishida, Publ. Astron. Soc. Jpn. 64, 7
(2012).

26. M.K. Kim, T. Hirota, M. Honma, H. Kobayashi, T. Bushimata,
Y.K. Choi, H.~Imai, K.~Iwadate, T. Jike, S. Kameno, et al., Publ.
Astron. Soc. Jpn. 60, 991 (2008).

27. V.S. Kislyuk, S.P. Rybka, A.I. Yatsenko, and N.V. Kharchenko,
Astron. Astrophys. 321, 660 (1997).

28. A.R. Klemola, R.B. Hanson, and B.F. Jones, in Galactic and
Solar System Optical Astrometry, Ed. L.V. Morrison and G.F.
Gilmore (Cambridge Univ. Press, Cambridge, 1994), p. 20.

29. J. Kovalevsky, L. Lindegren, M.A.C. Perryman, P.D. Hemenway,
K.J. Johnston, V.S.~Kislyuk, J.F. Lestrade, L.V. Morrison, et al.,
Astron. Astrophys. 323, 620 (1997).

30. T. Kurayama, T. Sasao, and H. Kobayashi, Astrophys. J. 627,
L49 (2005).

31. K. Kusuno, Y. Asaki, H. Imai, and T. Oyama, Astrophys. J. 774,
107 (2013).

32. F. van Leeuwen, Astron. Astrophys. 474, 653 (2007).

33. J.-F. Lestrade, R.A. Preston, D.L. Jones, R.B. Phillips,
A.E.E. Rogers, M.A. Titus, M.J.~Rioja, and D.C. Gabuzda, Astron.
Astrophys. 344, 1014 (1999).

34. L. Lindegren and J. Kovalevsky, Astron. Astrophys. 304, 189
(1995).

35. L. Loinard, R.M. Torres, A.J. Mioduszewski, L.F. Rodriguez,
R.A. Gonzalez-Lopezlira, R. Lachaume, V. Vazquez, and E. Gonzalez,
Astrophys. J. 671, 546 (2007).

36. M. L\'opez-Corredoira, H. Abedi, F. Garz\'on, and F. Figueras,
Astron. Astrophys. 126, 31 (2014).

37. C. Ma, E.F. Arias, T.M. Eubanks, A.L. Fey, A.-M. Gontier, C.S.
Jacobs, O.J. Sovers, B.A. Archinal, and P. Charlot, Astron. J.
116, 516 (1998).

38. C. Ma, F.E. Arias, G. Bianco, D.A. Boboltz, S.L. Bolotin, P.
Charlot, G. Engelhardt, A.L. Fey, R.A. Gaume, et al., VizieR
On-line Data Catalog: I/323 (2009).

39. K.M. Menten, M.J. Reid, J. Forbrich, and A. Brunthaler,
Astron. Astrophys. 474, 515 (2007).

40. J.C.A. Miller-Jones, G.R. Sivakoff, C. Knigge, E.G. Kording,
M. Templeton, and E.O. Waagen, Science 340, 950 (2013).

41. C. Min, N. Matsumoto, M.K. Kim, T. Hirota, K.M. Shibata, S.-H.
Cho, M.~Shizugami, and M.~Honma, arXiv:1401.5574 (2014).

42. A. Nakagawa, T. Omodaka, T. Handa, M. Honma, N. Kawaguchi, H.
Kobayashi, T.~Oyama, K. Sato, et al., arXiv:1404.4463 (2014).

43. A. Nakagawa, M. Tsushima, K. Ando, T. Bushimata, Y.K. Choi,
T.~Hirota, M.~Honma, H.~Imai, et al., Publ. Astron. Soc. Jpn. 60,
1013 (2008).

44. D. Nyu, A. Nakagawa, M. Matsui, H. Imai, Y. Sofue, T. Omodaka,
T. Kurayama, R.~Kamohara, T. Hirota, et al., Publ. Astron. Soc.
Jpn. 63, 63 (2011).

45. I. Platais, V. Kozhurina-Platais, T.M. Girard, W.F. van
Altena, C.E. Lopez, R.B. Hanson, A.R. Klemola, B.F. Jones, et al.,
Astron. Astrophys. 331, 1119 (1998a).

46. I. Platais, T.M. Girard, V. Kozhurina-Platais, W.F. van
Altena, C.E. Lopez, R.A. Mendez, W.-Z. Ma, T.-G. Yang, H.T.
MacGillivray, et al., Astron. J. 116, 2556 (1998b).

47. M.I. Ratner, N. Bartel, M.F. Bietenholz, D.E. Lebach, J.-F.
Lestrade, R.R. Ransom, and I.I. Shapiro, Astrophys. J. Suppl. Ser.
201, 5 (2012).

48. M.J. Reid, J.E. McClintock, R. Narayan, L. Gou, R.A.
Remillard, and J.A. Orosz, Astrophys. J. 742, 83 (2011).

49. M.J. Reid, K.M. Menten, A. Brunthaler, X.W. Zheng, T.M. Dame,
Y. Xu, Y.Wu, B.~Zhang, A. Sanna, M. Sato, et al., Astrophys. J.
783, 130 (2014).

50. S. Roeser, M. Demleitner, and E. Schilbach, Astrophys. J. 139,
2440 (2010).

51. S.P. Rybka and A.I. Yatsenko, Astron. Astrophys. Suppl. Ser.
121, 243 (1997).

52. K.L.J. Rygl, A. Brunthaler, M.J. Reid, K.M. Menten, H.J. van
Langevelde, and Y. Xu, Astron. Astrophys. 511, A2 (2010).

53. M.F. Skrutskie, R.M. Cutri, R. Stiening, M.D. Weinberg, S.
Schneider, J.M. Carpenter, C. Beichman, R. Capps, T. Chester, et
al., Astron. J. 131, 1163 (2006).

54. The HIPPARCOS and Tycho Catalogues, ESA SP--1200 (1997).

55. R.M. Torres, L. Loinard, A.J. Mioduszewski, A.F. Boden, R.
Franco-Hernandez, W.H.T. Vlemmings, and L.F. Rodriguez, Astrophys.
J. 747, 18 (2012).

56. R.M. Torres, L. Loinard, A.J. Mioduszewski, and L.F.
Rodriguez, Astrophys. J. 671, 1813 (2007).

57. H.-J. Tucholke, P. Brosche, and M. Odenkirchen, Astron.
Astrophys. Suppl. Ser. 124, 157 (1997).

58. W.H.T. Vlemmings, H.J. van Langevelde, P.J. Diamond, H.J.
Habing, and R.T. Schilizzi, Astron. Astrophys. 407, 213 (2003).

59. W.H.T. Vlemmings, and H.J. van Langevelde, Astron. Astrophys.
472, 547 (2007).

60. J. Vondr\'ak, C. Ron, and I. Pe\v sek, Astron. Astrophys. 319,
1020 (1997).

61. Z.-Y. Wu, J. Ma, and X. Zhou, Publ. Astron. Soc. Pacif. 123,
1313 (2011).

62. N. Zacharias, S.E. Urban, M.I. Zacharias, G. L. Wycoff, D.M.
Hall, M. E. Germain, E.R. Holdenried, and L. Winter, Astron. J.
127, 3043 (2004).

63. N. Zacharias, C. Finch, T. Girard, N. Hambly, G. Wycoff, M.I.
Zacharias, D.~Castillo, T.~Corbin, M. DiVittorio, S. Dutta, et
al., Astron. J. 139, 2184 (2009).

 64. B. Zhang, M.J. Reid, K.M. Menten, and X.W. Zheng, Astrophys. J. 744, 23 (2012).

65. Zi Zhu, Publ. Astron. Soc. Jpn. 53, L33 (2001).

 }

\end{document}